\documentclass[%
 reprint,
superscriptaddress,
 amsmath,amssymb,
 aps,
floatfix,
]{revtex4-2}

\usepackage[caption=false]{subfig}
\usepackage{graphicx}
\usepackage{dcolumn}
\usepackage{bm}
\usepackage[hidelinks]{hyperref}
\usepackage{xcolor}
\usepackage[normalem]{ulem}
\definecolor{ra}{rgb}{0.8, 0.0, 0.0}

\begin{document}

\preprint{APS/123-QED}
\title{Adding machine learning within Hamiltonians: Renormalization group transformations, symmetry breaking and restoration}
\author{Dimitrios Bachtis}
\email{dimitrios.bachtis@swansea.ac.uk}
\affiliation{Department of Mathematics,  Swansea University, Bay Campus, SA1 8EN, Swansea, Wales, UK}
\author{Gert Aarts}
\email{g.aarts@swansea.ac.uk}
\affiliation{Department of Physics, Swansea University, Singleton Campus, SA2 8PP, Swansea, Wales, UK}
\author{Biagio Lucini}
\email{b.lucini@swansea.ac.uk}
\affiliation{Department of Mathematics,  Swansea University, Bay Campus, SA1 8EN, Swansea, Wales, UK}%
\affiliation{Swansea Academy of Advanced Computing, Swansea University, Bay Campus, SA1 8EN, Swansea, Wales, UK}

\include{ms.bib}

\date{September 30, 2020}

\begin{abstract}

We present a physical interpretation of machine learning functions, opening up the possibility to control properties of statistical systems via the inclusion of these functions in Hamiltonians. In particular, we include the predictive function of a neural network, designed for phase classification, as a conjugate variable coupled to an external field within the Hamiltonian of a system. Results in the two-dimensional Ising model evidence that the field can induce an order-disorder phase transition  by breaking or restoring the symmetry, in contrast with the field of the conventional order parameter which causes explicit symmetry breaking. The critical behaviour is then studied by proposing a Hamiltonian-agnostic reweighting approach and forming a renormalization group mapping on quantities derived from the neural network. Accurate estimates of the critical point and of the critical exponents related to the operators that govern the divergence of the correlation length are provided.  We conclude by discussing how the method provides an essential step towards bridging machine learning and physics.

\end{abstract}

\maketitle

\section{Introduction}

At the heart of our understanding of phase transitions lies a mathematical apparatus called the renormalization group \citep{PhysRevB.4.3174,PhysRevB.4.3184,PhysicsPhysiqueFizika.2.263,RevModPhys.47.773,PhysRevLett.28.240,WILSON197475}. Central concepts behind its application are those of scale invariance and universality: the former relates to the observation that at criticality the considered phenomena can be described by a scale-invariant theory, and the latter to the notion that systems seemingly unrelated in their microscopic descriptions have a large-scale behaviour that is governed by an identical set of relevant operators. Computational frameworks of the renormalization group \citep{PhysRevLett.42.859,PhysRevLett.37.461} have seen resounding success in condensed matter systems \citep{PhysRevB.30.3866,PhysRevLett.89.275701,PhysRevLett.76.2613,PhysRevE.95.053305} and lattice field theories \citep{PhysRevLett.71.3063,PhysRevD.80.034505,PhysRevD.82.014506}. 

Recently, deep learning \citep{GoodBengCour16}, which pertains to a class of machine learning methods that progressively extract hierarchical structures in data, has impacted certain aspects of computational science. Artificial neural networks, consisting of multiple layers, have been efficiently applied in various research fields, including particle physics and cosmology as well as statistical mechanics. For a recent review see Refs.~\citep{Carleo_2019,Carrasquilla_2020review}. Notable examples include the use of neural networks to study phase transitions \citep{Carrasquilla2017,vanNieuwenburg2017,PhysRevLett.120.257204,Rodriguez-Nieva2019,GIANNETTI2019114639,chernodub2020topological,boyda2020machinelearning} and quantum many-body systems \citep{Carleo_2017manybody}, while investigations at the intersection of machine learning and the renormalization group have recently emerged \citep{PhysRevLett.121.260601,mehta2014exact,Koch-Januszrg,PhysRevE.97.053304,bny2013deep}. In consequence, an in-depth understanding of the underlying mechanics of machine learning algorithms, as well as a simultaneous advancement of efficient ways to implement them in physical problems, is a crucial step to be undertaken by physicists \citep{Zdeborova2020ml}. 

In this paper a physical interpretation of machine learning is presented. In particular, we consider the predictive function of a neural network, designed for phase classification, as a conjugate variable coupled to an external field, and introduce it as a term in the Hamiltonian of a system. Given this formulation, we propose reweighting that is agnostic to the original Hamiltonian to explore if the external field generates a richer structure than the one associated with the conventional order parameter, and if it can induce a phase transition by breaking or restoring the system's symmetry. The critical behaviour can then be investigated using histogram-reweighted extrapolations from configurations obtained in one phase and without knowledge of the Hamiltonian.

To study the phase transition, we propose a real-space renormalization group transformation that is formulated on machine learning quantities. In particular, a mapping is established between an original and a rescaled system using the neural network function and its  field, overcoming the need to rely on observables related to the original Hamiltonian. We then explore, based on minimally-sized lattices, its capability to locate the critical fixed point and to extract the operators of the renormalization group transformation. The entirety of critical exponents can then be obtained using scaling relations and a complete study of the phase transition can be conducted.

We validate our proposal in the two-dimensional Ising model using quantities derived from the machine learning algorithm. By giving a physical interpretation to the function of a neural network as a Hamiltonian term, we explore the effect of the coupled field on the considered system, demonstrate that it can break or restore the reflection symmetry by inducing a phase transition, and extract with high accuracy the location of the critical inverse temperature and the operators of the renormalization group transformation that govern the divergence of the correlation length.

\begin{figure*}
\includegraphics[width=16.2cm]{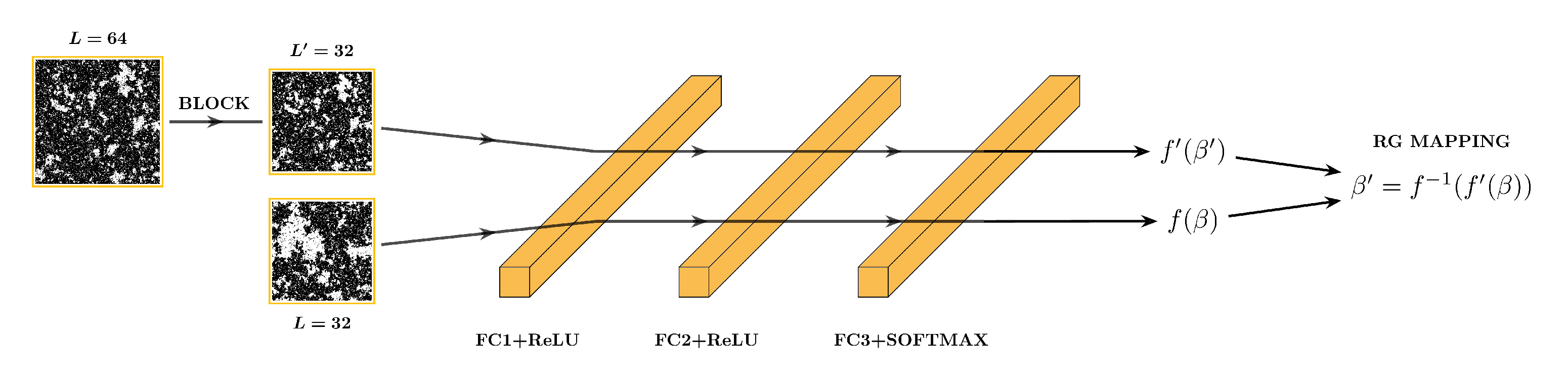}
\caption{\label{fig:conv2d}The architecture of the fully-connected neural network. A renormalization group mapping is established between an original and a rescaled system based on the predictive function of the neural network. }
\end{figure*}

\section{Neural Networks as Hamiltonian Terms}

We consider a statistical system, such as the Ising model (see Appendix~\ref{app:ising}), which is described by a Hamiltonian $E$. The equilibrium occupation probabilities of the system are of Boltzmann form and are given by:
\begin{equation} \label{probs}
p_{\sigma}= \frac{\exp[-\beta E_{\sigma}]}{\sum_{\sigma} \exp[-\beta E_{\sigma}]},
\end{equation}
where $\beta$ is the inverse temperature, $\sigma$ a state of the system and $Z=\sum_{\sigma} \exp[-\beta E_{\sigma}]$ the partition function.  When the system is in equilibrium the expectation value of an arbitrary observable $O$ is:
\begin{equation} \label{estim}
\langle O \rangle=\frac{\sum_{\sigma} {O_{\sigma} \exp[{-\beta E_{\sigma}]}}}{\sum_{\sigma}  \exp[{-\beta E_{\sigma}]}}.
\end{equation}

After a neural network is trained on a system for phase classification (see Appendix~\ref{app:fcn} and Fig.~\ref{fig:conv2d}), the learned neural network function $f(\cdot)$ can be applied to a configuration $\sigma$, converting $f_{\sigma}$ into a statistical mechanical observable with an associated Boltzmann weight \citep{bachtis2020extending}. In addition, we consider $f_{\sigma}$ as equivalent to the conditional probability $f_{\sigma} \equiv P_{\sigma}^{b}$ that a configuration belongs in the broken-symmetry phase. Consequently, $f_{\sigma}$  is an intensive quantity bound between $[0,1]$ and since it has no dependence on the size of the system we can multiply it with the volume $V$ and recast $Vf_{\sigma}$ as an extensive property. 

 We are now able to investigate the extensive neural network function $Vf$ by introducing it in the Hamiltonian of the system. Fields that interact with a system have conjugate variables which represent the response of the system to the perturbation of the corresponding field. We therefore consider $Vf$ as a conjugate variable that couples to an external field $Y$ and define a modified Hamiltonian:
\begin{equation}\label{eq:Hmod}
E_{Y}= E-VfY.
\end{equation}

The expectation value of the neural network function can then be expressed as a derivative of the modified partition function $Z_{Y}$ in terms of the field:
\begin{equation}\label{eq:fmod}
\langle f \rangle = \frac{1}{\beta V} \frac{\partial \ln Z_{Y}}{ \partial Y}=\frac{\sum_{\sigma} {f_{\sigma} \exp[{-\beta E_{\sigma} +\beta V f_{\sigma}Y]}}}{\sum_{\sigma}  \exp[{-\beta E_{\sigma}+\beta V f_{\sigma}Y]}}.
\end{equation}

Setting the neural network field $Y$ to zero results in the standard definition of Eq.~(\ref{estim}). Nevertheless, a derivation of Eq.~(\ref{eq:fmod}) in terms of the field gives:
\begin{equation}
\chi_{f}=\frac{\partial \langle f \rangle}{\partial Y} = \beta V (\langle f^{2} \rangle - \langle f \rangle^{2}).
\end{equation}

The quantity $\chi_{f}$ is recognized as a susceptibility. It is a measure of the response of the predictive function $f$ to changes in the neural network field $Y$. Consequently, the opportunity to study the effect of a non-zero field $Y \neq 0$ in the statistical system is now available. One way to achieve this is to conduct Monte Carlo sampling using the modified Hamiltonian of Eq.~(\ref{eq:Hmod}) to obtain configurations of this modified system. However, an alternative option that overcomes the need for sampling is the use of histogram reweighting \citep{PhysRevLett.61.2635,PhysRevLett.63.1195}, where machine learning derived observables can also be reweighted in parameter space \citep{bachtis2020extending}. 

\section{Symmetry Breaking and Restoration}

Consider a set of $N$ obtained configurations $\sigma_{i}$ from a system whose explicit form of the Hamiltonian $E$ is not known. These configurations have been drawn from an equilibrium distribution, described by Eq.~(\ref{probs}), and can be utilized with reweighting to predict the behaviour of the modified system, when the neural network field $Y$ is set to non zero-values. 

To achieve this we define the expectation value for an arbitrary observable $O$, estimated during a Markov chain Monte Carlo simulation, in the modified system that we aim to sample:
\begin{equation} \label{estimo}
\langle O \rangle=\frac{\sum_{i=1}^{N} O_{\sigma_{i}} \tilde{p}_{\sigma_i}^{-1}  \exp[-\beta E_{\sigma_{i}} + \beta V f_{\sigma_{i}} Y]}{\sum_{i=1}^{N} \tilde{p}_{\sigma_i}^{-1}  \exp[-\beta E_{\sigma_{i}} + \beta V f_{\sigma_{i}} Y]},
\end{equation}
where $\tilde{p}$ are the sampling probabilities of the equilibrium distribution. The probabilities $p$ of the original system,  defined in Eq.~(\ref{probs}),  can be substituted for $\tilde{p}_{\sigma_{i}}$ to obtain:
\begin{equation} \label{eq:reweight}
\langle O \rangle=\frac{\sum_{i=1}^{N} O_{\sigma_{i}} \exp[ \beta V f_{\sigma_{i}} Y]}{\sum_{i=1}^{N} \exp[ \beta V f_{\sigma_{i}} Y]},
\end{equation}

\begin{figure}[t]
\includegraphics[width=8.6cm]{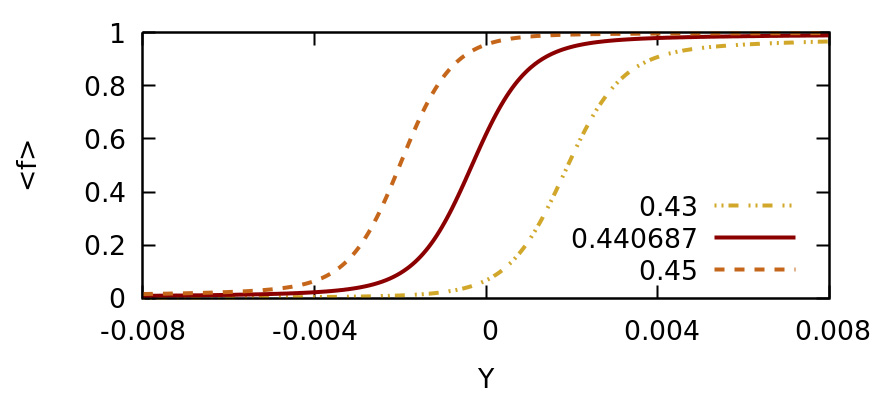}
\caption{\label{fig:field1} Mean neural network function $\langle f \rangle$ versus external field  $Y$ at inverse temperature $\beta=0.43,0,440687,0.45$ (right to left). The statistical uncertainty is comparable with the width of the lines. }
\end{figure}

Using Eq.~(\ref{eq:reweight}) one can calculate the expectation value of an observable for a modified system with a non-zero neural network field $Y \neq 0$ by using configurations of the original system sampled at inverse temperature $\beta$ and with zero field $Y=0$. Before investigating the effect of $Y \neq 0$  to machine learning devised observables we recall that the function $f$ has emerged by training the neural network on obtained configurations, where no knowledge about the explicit form of the Hamiltonian is introduced. As a result, both Eq.~(\ref{eq:reweight}) and the function $f$ have no immediate dependence on the Hamiltonian and the opportunity to conduct reweighting that is Hamiltonian-agnostic is present.

 For the case of the neural network function $f$, results that have been obtained using Eq.~(\ref{eq:reweight}) can be seen in Fig.~\ref{fig:field1}. A two-dimensional Ising model of size $L=64$ at each dimension is simulated at inverse temperatures $\beta=0.43$ in the symmetric phase, $\beta=0.440687$ in the known inverse critical temperature and $\beta=0.45$ in the broken-symmetry phase. We observe that regardless of the phase that the system is in, positive and negative values of the external field $Y$ drive the system towards the broken-symmetry or the symmetric phase, respectively. To gain further insights, we recall that the neural network function is correlated with the probability $P^{s}$ that a configuration is associated with the symmetric phase through $f \equiv P^{b}=1-P^{s}$. Consequently, the associated field which is coupled to $f$ is anticipated to have the observed behaviour.

\begin{figure}[t]
\includegraphics[width=8.6cm]{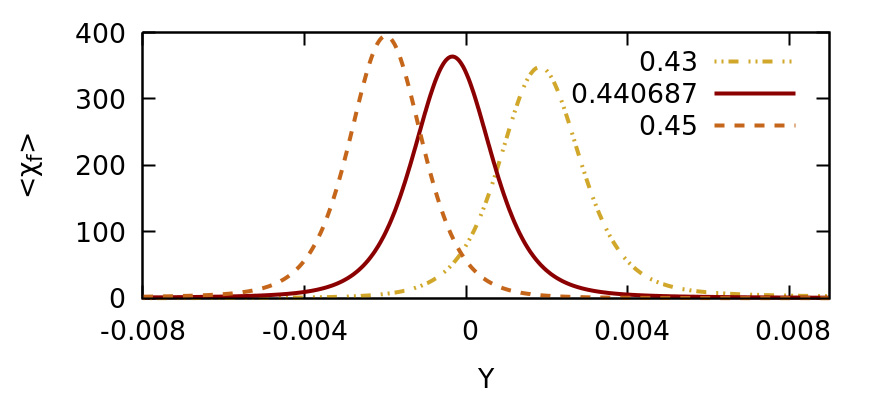}
\caption{\label{fig:field2} Mean susceptibility of the neural network function $\langle \chi_{f} \rangle$ versus external field $Y$ at inverse temperature $\beta=0.43,0,440687,0.45$ (right to left).  The statistical uncertainty is comparable with the width of the lines.}
\end{figure}

To measure the response of the predictive function $f$ to changes in the field we calculate the susceptibility $\chi_{f}$, which is depicted in Fig.~\ref{fig:field2}. We note that $\chi_{f}$ has maximum values, evidencing the crossing of a phase transition. The results indicate that the field can induce an order-disorder phase transition in the Ising model by breaking or restoring the symmetry of the system. This is in contrast with the external field associated with the conventional order parameter (the magnetization) that, irrespective of its sign, induces an explicit breaking of the symmetry. Based on universality, the emerging phase transition is assumed to be governed by the same relevant operators and the critical behaviour of the neural network field can now be studied with the renormalization group.

\section{Renormalization group: Locating the critical fixed point}

\begin{figure}[b]
\includegraphics[width=8.6cm]{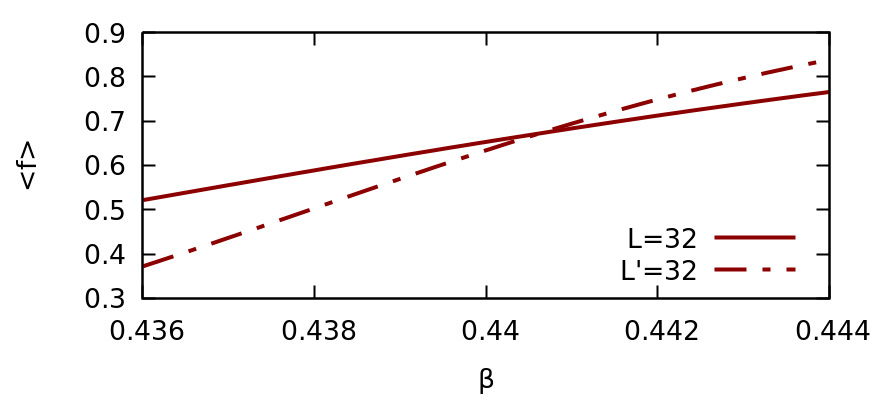}
\caption{\label{fig:prob} Mean neural network function $\langle f \rangle$ versus inverse temperature $\beta$ for an original and a rescaled system of size $L=L'=32$. }
\end{figure}

We consider a configuration of an Ising model that has been obtained at a specific inverse temperature $\beta$ and has an emerged correlation length $\xi$.  By applying the blocking procedure with the majority rule and a rescaling factor of $b=2$ (see Appendix~\ref{app:block}), the reduction in the lattice size $L'=L/b$ will also induce an analogous reduction in the given correlation length:
\begin{equation}\label{eq:xiprime}
\xi'=\frac{\xi}{b}.
\end{equation}
 
The correlation length $\xi$ is a quantity that emerges dynamically when the system is approaching the critical point $\beta \approx \beta_{c}$ and is therefore dependent on the value of the inverse temperature $\xi(\beta)$. The rescaled system has a reduced correlation length $\xi'$ and is, consequently, representative of an Ising model at a different inverse temperature, with $\xi'(\beta')$. At the critical fixed point $\beta'=\beta=\beta_{c}$  the correlation length in the thermodynamic limit $\xi(\beta_{c}, L=\infty)$ diverges, and intensive quantities of the original and the rescaled system become equal.

This opens up the opportunity to use an observable derived from a machine learning algorithm to locate the critical fixed point. In particular, we consider at $\beta = \beta_{c}$, the neural network function $f$ which has been expressed as a statistical mechanical observable and is therefore dependent on the inverse temperature:
\begin{equation} \label{eq:fcrit}
f(\beta_{c})=f'(\beta_{c}).
\end{equation}

In Fig.~\ref{fig:prob}, the predictive function $f$ has been drawn for an original and a rescaled system. We recall that, under the assumption that configurations of the rescaled system appear with the Boltzmann probabilities of the original Ising Hamiltonian, observables of the rescaled system can be reweighted as observables of the original \citep{newmanb99}.  We note that the two lines cross at $\beta_{c}^{f} \approx 0.44055$, yielding a first estimate of the location of the critical inverse temperature. The results are obtained using reweighting on a Monte Carlo dataset which has been simulated near the known inverse critical temperature $\beta_{c} \approx 0.440687$.  When the inverse critical temperature isn't known it can be estimated by iterating the same procedure until convergence  \citep{PhysRevLett.42.859,newmanb99}.  Given this knowledge,  the operators and the critical fixed point  of the renormalization group transformation can then be calculated in a quantitative manner.

\begin{figure}[t]
\includegraphics[width=8.6cm]{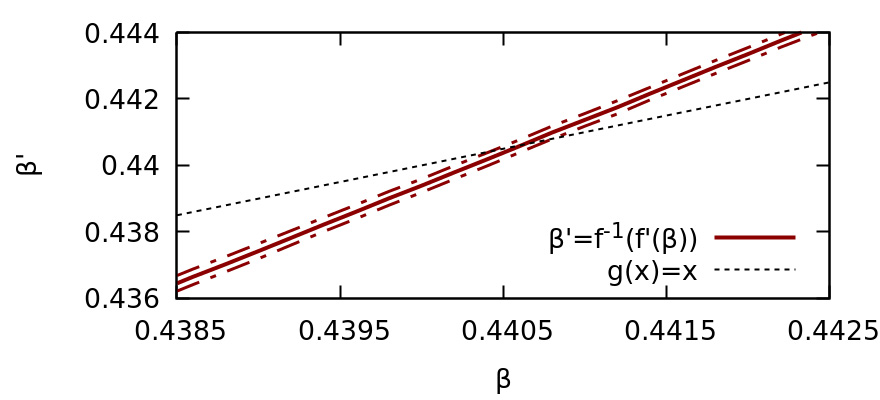}
\caption{\label{fig:mapb} Rescaled inverse temperature $\beta'$ versus inverse temperature $\beta$. The intersection with $g(x)=x$ gives the critical fixed point $\beta=\beta'=\beta_{c}$. The dashed lines adjacent to the solid line indicate the statistical uncertainty.}
\end{figure}

\section{Extracting Operators of the Renormalization Group}

The original and the rescaled systems are located at inverse temperatures $\beta$ and $\beta'$, and their neural network functions are, therefore, related through:
\begin{equation}\label{eq:fmapc}
f(\beta')=f'(\beta).
\end{equation}

For the case of the inverse critical temperature, Eq.~(\ref{eq:fmapc}) reduces to Eq.~(\ref{eq:fcrit}). We are now able to form, based on Eq.~(\ref{eq:fmapc}), a renormalization group mapping that associates the two inverse temperatures (see Fig.~\ref{fig:conv2d}):
\begin{equation} \label{eq:mapping}
\beta'=f^{-1}(f'(\beta)).
\end{equation}

The correlation length then diverges in the thermodynamic limit according to relations $\xi \sim |t|^{-\nu}$ and  $\xi' \sim |t'|^{-\nu}$ for an original and a rescaled system, respectively, where $ t=(\beta_{c}-\beta)/\beta_{c}$ is the reduced inverse temperature. Dividing the two equations,  c.f.~Eq.~(\ref{eq:xiprime}), we obtain:
\begin{equation} \label{eq:tovert}
\bigg(\frac{t}{t'}\bigg)^{-\nu}=b.
\end{equation}

The renormalization group mapping is then linearized through a Taylor expansion to leading order in the proximity of the fixed point \citep{dombg}, to obtain:
\begin{equation}
\beta_{c}-\beta'= (\beta_{c}-\beta) \frac{d\beta'}{d\beta} \bigg|_{\beta_{c}}.
\end{equation}

By substituting into Eq.~(\ref{eq:tovert}) we are able to calculate the correlation length exponent:
\begin{equation}\label{eq:correxp}
\nu= \frac{\ln b}{\ln \frac{d\beta'}{d\beta} \Big|_{\beta_{c}}}.
\end{equation}

In Fig.~\ref{fig:mapb} results based on Eq.~(\ref{eq:mapping}) are depicted using Hamiltonian-dependent reweighting on the inverse temperature \citep{bachtis2020extending} . We obtain an estimate of the critical fixed point $\beta_c=0.44063(21)$ and the correlation length exponent $\nu=1.01(2)$.

Since the neural network field $Y$ induces a phase transition in the system it is bound to affect the correlation length. Another critical exponent $\theta_{Y}$ can then be defined when the field converges to zero at the critical fixed point:
\begin{equation}
\xi \sim |Y|^{-\theta_{Y}}.
\end{equation}
 \begin{figure}[t]
\includegraphics[width=8.6cm]{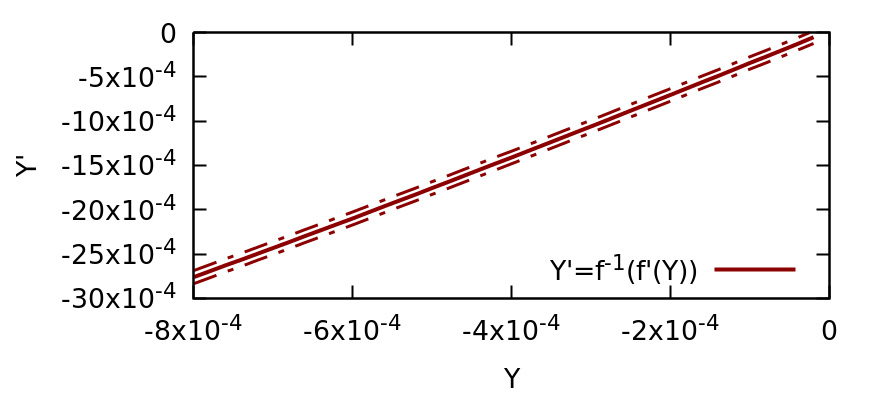}
\caption{\label{fig:mapf} Rescaled field $Y'$ versus original field $Y$ at $\beta=0.440687$. The dashed lines indicate the statistical uncertainty. }
\end{figure}

Following an analogous derivation, and formulating a mapping $Y'=f^{-1}(f'(Y))$ for the field, the corresponding critical exponent is then calculated through:
\begin{equation}\label{eq:fieldexp}
\theta_{Y}= \frac{\ln b}{\ln \frac{d Y'}{d Y} \Big|_{Y=0}}.
\end{equation}

Fig.~\ref{fig:mapf} shows the results for the case of the neural network field, where we obtain the value of the critical exponent $\theta_{Y}=0.534(3)$, using Hamiltonian-agnostic reweighting based on Eq.~(\ref{eq:reweight}).

 The phase transition of the Ising model is described in completeness based on two relevant operators $\nu$ and $\theta$. The exponent $\theta$ governs the divergence of the correlation length in terms of the external field $h$ that is coupled to the conventional order parameter. We note that the predictive function $f$ is reminiscent of an effective order parameter (see Fig.~\ref{fig:field1}). We find that the numerical value of the exponent $\theta_{Y}$ agrees within statistical errors with $\theta=8/15$. We hence conclude that $Y$ couples to the same relevant operator as the external magnetic field. The results are summarized in Table~\ref{tab:table2} and the remaining critical exponents can be calculated through scaling relations (see Appendix~\ref{app:ising}). 
 
 We emphasize that the operators and the critical fixed point have been calculated using observables derived from the neural network implementation and their reweighted extrapolations where no explicit information about the symmetries of the Hamiltonian was introduced.

\begin{table}[b]
\caption{\label{tab:table2}
Estimates for the critical exponents $\nu$, $\theta_{Y}$ and the critical fixed point $\beta_{c}$ of the two-dimensional Ising model.}
\begin{ruledtabular}
\begin{tabular}{ccccc}
 &$\beta_{c}$&$\nu$ &$ \theta_{Y}, \theta$ \\
\hline
RG+NN & 0.44063(21) & 1.01(2) & $\theta_{Y}=0.534(3)$ \\
  Exact & $\ln(1+\sqrt{2})/2$ & $1$ & $\theta=8/15$  
\end{tabular}
\end{ruledtabular}
\end{table}

\section{Conclusions}

 The inclusion of the predictive function in the Hamiltonian enables the calculation of a relevant operator, namely the magnetic field exponent $\theta$, that was previously inaccessible through supervised machine learning methods which are agnostic to the symmetries of the system. The application of a renormalization group transformation diminishes finite size effects \citep{newmanb99}, and a highly accurate calculation of the critical fixed point and the relevant operators of the two-dimensional Ising model is conducted on minimally-sized lattices. The results, obtained by one iteration of a spin blocking transformation, are comparable with traditional renormalization group techniques \citep{PhysRevLett.42.859}, with the added benefit that the method is agnostic to the Hamiltonian of the system and can therefore be implemented in cases where an order parameter is absent or unknown \citep{Carrasquilla2017}. When knowledge of the Hamiltonian is included in the calculations, the possibility to investigate the contribution of the introduced machine learning term in the calculation of critical exponents within the framework of the Monte Carlo renormalization group \citep{PhysRevLett.42.859} additionally exists. 
 
 Furthermore, the method extends reweighting, a technique that is applicable to a wide range of ensembles \citep{PhysRevLett.61.2635}, by introducing a novel Hamiltonian-agnostic approach to extrapolate machine learning quantities in parameter space without requiring any knowledge about the energy of the system. Predictive functions have been successfully constructed in cases of first and second-order phase transitions for spin models and quantum field theories \citep{bachtis2020mapping}. As the proposed method only requires a predictive function, and no knowledge about the Hamiltonian (see Eq.~\ref{eq:reweight}),  there exists no a priori argument that forbids the method in being applied to a wide range of systems, across different ensembles. It is therefore anticipated to be applicable in phase transitions of systems simulated in ensembles such as the canonical, grand-canonical, isothermal-isobaric and quantum Monte Carlo simulations across systems in statistical mechanics, condensed matter physics and lattice field theories.

Machine learning can become physically interpretable by being introduced as a term in the Hamiltonian and numerous research directions can be anticipated. Any machine learning function can, in principle, be instilled within Hamiltonians to control properties of a system, such as to induce symmetry breaking or symmetry restoration. The possibility to include a function learned from a simple model to study a complicated one exists \citep{bachtis2020mapping}. Most importantly, by using Monte Carlo simulations to sample configurations of modified systems that include neural network functions, an in-depth understanding of the underlying mechanics of machine learning can be obtained. 

In conclusion, by including machine learning as a term in the Hamiltonian an essential step towards bridging machine learning and physics is established, one that could potentially alter our understanding of machine learning algorithms and their effects on systems.

\section{\label{sec:level5}Acknowledgements}
The authors received funding from the European Research Council (ERC) under the European Union's Horizon 2020 research and innovation programme under grant agreement No 813942. The work of GA and BL has been supported in part by the UKRI Science and Technology Facilities Council (STFC) Consolidated Grant ST/P00055X/1. The work of BL is further supported in part by the Royal Society Wolfson Research Merit Award WM170010 and by the Leverhulme Foundation Research Fellowship RF-2020-461\textbackslash 9. Numerical simulations have been performed on the Swansea SUNBIRD system. This  system is part of the Supercomputing Wales project, which is part-funded by the European Regional Development Fund (ERDF) via Welsh Government. We thank COST Action CA15213 THOR for support.

\appendix

\section{The Ising Model \label{app:ising}}

We consider the two-dimensional Ising model on a square lattice which is described by the Hamiltonian:
\begin{equation}
E=-J\sum_{\langle ij \rangle} s_{i} s_{j}-h\sum_{i} s_{i},
\end{equation}
where $\langle ij \rangle$ is a sum over nearest neighbor interactions, $J$ is the coupling constant which is set to one and $h$ the external magnetic field which is set to zero. The system undergoes a second-order phase transition at the critical inverse temperature $\beta_{c}$:
\begin{equation}
\beta_{c}= \frac{1}{2} \ln(1+\sqrt{2}) \approx 0.440687.
\end{equation}

The order parameter of the Ising model is the magnetization. We often consider the absolute magnetization, normalized by the volume $V=L \times L$ of the system:
\begin{equation}
m=\frac{1}{V} \Big | \sum_{i} s_{i} \Big|.
\end{equation}

The fluctuations of the magnetization are equivalent to the magnetic susceptibility, defined as:

\begin{equation}
\chi= \beta V (\langle m^{2} \rangle- \langle m \rangle^{2}).
\end{equation}

To measure the distance from the critical point a dimensionless reduced inverse temperature is defined:
\begin{equation}
t= \frac{\beta_{c}-\beta}{\beta_{c}}.
\end{equation}

As the system approaches the critical temperature $\beta \approx \beta_{c}$, finite size effects dominate and fluctuations, such as the magnetic susceptibility $\chi$, have maximum values which act as phase transition indicators. 

The phase transition of the Ising model is described in completeness by two relevant operators that govern the divergence of the correlation length. One is the critical exponent $\nu$ which is defined via:
\begin{equation}
\xi \sim |t|^{-\nu},
\end{equation}
and the second one is the critical exponent $\theta$ which is given through:
\begin{equation}\label{eq:apptheta}
\xi \sim |h|^{-\theta},
\end{equation}
where $h$ is the external magnetic field. Eq.~(\ref{eq:apptheta}) is valid when $\beta=\beta_{c}$ and $h \rightarrow 0$.

 Given the two relevant operators $\nu$ and $\theta$, the critical exponents that govern the divergence of the specific heat ($\alpha$), magnetization ($\beta^{(m)}$) and magnetic susceptibility ($\gamma$) can then be calculated through scaling relations:
\begin{eqnarray}
\alpha  =& 2-\nu d \\
\beta^{(m)} =& \nu \Big[ d - \frac{1}{\theta} \Big] \\
\gamma  =&  \nu \Big[ \frac{2}{\theta} - d \Big] \\
\delta  =& \frac{1}{d \theta-1},
\end{eqnarray}
where $d=2$ is the dimensionality of the system.

\section{ Neural Network Architecture and Simulation details \label{app:fcn}}

The neural network architecture is comprised of a fully-connected layer (FC1) with a rectified linear unit (ReLU) non-linear function, defined as $k(x)=\max(0,x)$. The result is then passed to a second fully-connected layer (FC2) with $32$ units and a ReLU function, and is subsequently forwarded to a third fully-connected (FC3) with $2$ units and a softmax function. The training is conducted based on the Adam algorithm with a learning rate of $10^{-4}$ and a batch size of $8$. The architecture is implemented with TensorFlow and the Keras library.

\begin{figure}[t]
\includegraphics[width=7cm]{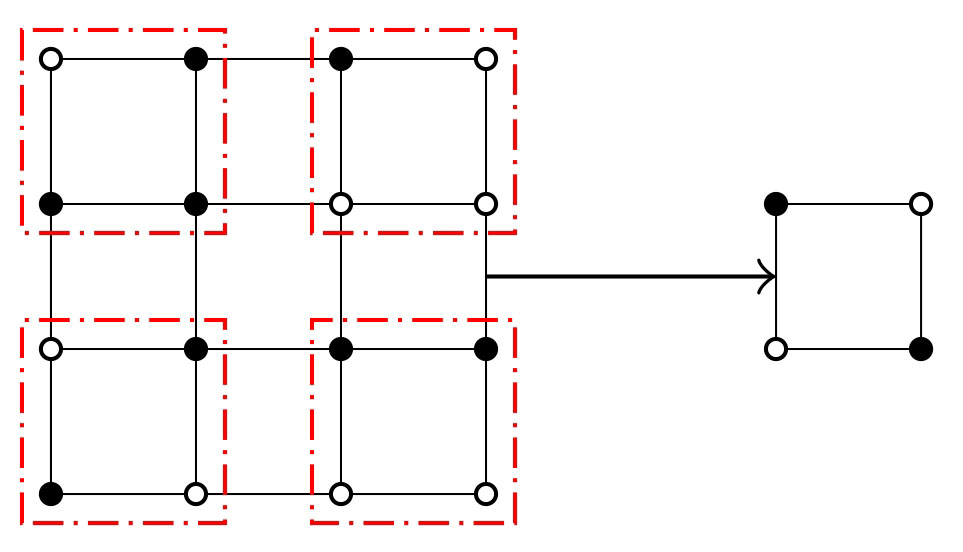}
\caption{\label{fig:blocking} An illustration of the blocking procedure with a rescaling factor of $b=2$ and the majority rule. For the two cases at the bottom, the choice of the rescaled degree of freedom was made arbitrarily.}
\end{figure}

Configurations are sampled with Markov chain Monte Carlo simulations using the Wolff algorithm \citep{PhysRevLett.62.361}, and are chosen to be minimally correlated. The data set is comprised of $1000$ configurations per each inverse temperature, where $100$ have been chosen to create a validation set. Specifically the training range chosen to sample configurations is $\beta=0.27,\ldots,0.36$ in the symmetric phase and $\beta=0.52,\ldots,0.61$ in the broken-symmetry phase with a step size of $0.01$. The architecture has been optimized on the Ising model based on the values of the training and the validation loss.

\section{The Blocking Procedure\label{app:block}}

A common choice of a transformation for the Ising model is the blocking procedure with a rescaling factor of $b=2$  and the majority rule. To apply the blocking procedure the lattice structure is initially separated into blocks of size $b \times b$. Within each block of the original system, degrees of freedom that have distinct values are counted and a majority rule defines the rescaled degree of freedom. When the counted degrees are equal, the choice is made arbitrarily. For an illustration of all possible outcomes see Fig.~\ref{fig:blocking}. The application of a blocking procedure preserves the large-scale information of a configuration and we assume that it results in a rescaled system of size $L'=L/b$ that is a valid representation of an Ising model \citep{newmanb99}. 

\section{Binning Error Analysis\label{app:binning}}

The error analysis is conducted with the binning method to address statistical errors associated with the finite Monte Carlo data sets. In particular, each Monte Carlo data set, comprised of $10000$ minimally correlated configurations is separated into $n=10$ groups. Results are calculated using the data in each group. The standard deviation is then given by:

\begin{equation}
\sigma_{x} = \sqrt{\frac{1}{n-1} (\overline{x^{2}} - \overline{x}^{2}}).
\end{equation}

\bibliography{ms}
\end{document}